\documentclass[aps,prl,10pt,twocolumn,bibnotes,superscriptaddress]{revtex4-2}

\usepackage[english]{babel}
\usepackage[utf8]{inputenc}
\usepackage{amssymb}
\usepackage{amsmath}
\usepackage{xcolor}
\usepackage{wasysym}
\usepackage{graphicx}
\usepackage{multirow}
\usepackage[T1]{fontenc}
\usepackage[pdftex, pdftitle={Article}, pdfauthor={Author}]{hyperref}
\definecolor{darkgreen}{rgb}{0.0, 0.5, 0.0}

\usepackage{pgffor}
\usepackage{pdfpages}
\usepackage{etoolbox}

\makeatletter
\patchcmd{\@outputpage@head}{\@ifx{\LS@rot\@undefined}{}{\LS@rot}}{}{}{}
\makeatother

\begin{document}

\title{Deep learning of quantum entanglement from incomplete measurements}

\author{Dominik Koutn\'{y}}
\affiliation{Department of Optics, Palack\'{y} University, 17. listopadu 12, 77146 Olomouc, Czech Republic}

\author{Laia Ginés}
\affiliation{Department of Physics, Stockholm University, 10691 Stockholm, Sweden}

\author{Magdalena Moczała-Dusanowska}
\affiliation{Princeton Institute of Materials, Princeton University, Princeton, NJ 08544, USA}

\author{Sven Höfling}
\affiliation{Technische Physik, Physikalisches Institut and W\"urzburg-Dresden Cluster of Excellence ct.qmat, Universit\"at W\"urzburg, Am Hubland, D-97074 W\"urzburg, Germany}

\author{Christian Schneider}
\affiliation{Institut of Physics, University of Oldenburg, D-26129 Oldenburg, Germany}

\author{Ana Predojevi\'c}
\affiliation{Department of Physics, Stockholm University, 10691 Stockholm, Sweden}

\author{Miroslav Je\v{z}ek}
\email[Correspondence email address: ]{jezek@optics.upol.cz}
\affiliation{Department of Optics, Palack\'{y} University, 17. listopadu 12, 77146 Olomouc, Czech Republic}


\begin{abstract}
The quantification of the entanglement present in a physical system is of para\-mount importance for fundamental research and many cutting-edge applications. Currently, achieving this goal requires either a priori knowledge on the system or very demanding experimental procedures such as full state tomography or collective measurements. Here, we demonstrate that by employing neural networks we can quantify the degree of entanglement without needing to know the full description of the quantum state. Our method allows for direct quantification of the quantum correlations using an incomplete set of local measurements. Despite using undersampled measurements, we achieve a quantification error of up to an order of magnitude lower than the state-of-the-art quantum tomography.
Furthermore, we achieve this result employing networks trained using exclusively simulated data. Finally, we derive a method based on a convolutional network input that can accept data from various measurement scenarios and perform, to some extent, independently of the measurement device.
\end{abstract}

\maketitle

\section{Introduction}
Physical measurements performed on individual parties of an entangled system reveal strong correlations \cite{Horodecki2009}, which give rise to nonclassical and nonlocal effects \cite{EPR1935,Bell1964}. Aforesaid effects are the essential element of fundamental tests of quantum mechanics, including direct experimental verification of quantum nonlocality \cite{Hensen2015,Shalm2015,Giustina2015}. The critical role of entanglement was demonstrated also on the opposite scale of the complexity spectra in macroscopic phase transitions \cite{Osterloh2002,Osborne2002,Amico2008}. Besides the fundamental aspects, entanglement is an essential tool for quantum information processing and it allows for reaching the quantum advantage \cite{Zhong2020,Madsen2022}. Modern quantum communication networks rely crucially on entanglement sources \cite{Ursin2018,Pan2020,Laurat2020,Trotta2021}. Consequently, the characterization of entanglement is paramount for both fundamental research and quantum applications \cite{Erhard2020,Eisert2020}.

Here, we adopt methods of deep learning to tackle the long-standing problem of efficient and accurate entanglement quantification. Our approach determines the degree of entanglement of a generic quantum state {\em directly} from an arbitrary set of local measurements. Despite the deep learning models being trained on simulated measurements, they excel when applied to real-world measurement data. We quantify photonic entanglement generated by two distinct systems: a nonlinear parametric process and a semiconductor quantum dot.

Reliable entanglement quantification represents an open problem in quantum physics. Direct measurement of entanglement can be achieved by exploiting quantum interference of two (or more) identical copies of a physical system \cite{Horodecki2003,Fiurasek2004,Walborn2006,Islam2015,Kaufman2016}. This multi-copy approach roots in measuring non-linear functions of quantum states \cite{Filip2002,Ekert2002}. However, such  measurements are experimentally highly demanding, which has spurred the research of single-copy entanglement detection utilizing only local measurements, such as quantum tomography.

Quantum tomography provides the full description of a quantum state including the degree of entanglement \cite{VogelRisken1989,Paris2004Springer}. However, the total number of measurements required for quantum tomography increases exponentially with the number of qubits or quantum degrees of freedom, which renders the approach inherently not scalable \cite{Kwiat2005,Blatt2011,Pan2018}. Several methods have been developed to make this scaling more favorable, nevertheless, by imposing an a priori structure or symmetry to the system \cite{Gross2010,Cramer2010,Toth2010,Lanyon2017}.
When a few-parameter model of quantum state is assumed, quantum estimation can be used for optimal inferring of the state entanglement \cite{Brida2010,Brida2011,Benedetti2013}.
Another approach to emulate quantum correlations \cite{Deng2017} with fewer resources relies on neural-network quantum states \cite{Carleo2017,Torlai2018,Carleo2018,Hartmann2019,Garttner2021}. Yet, this method suffers from the sign problem, solving of which requires further assumptions about the state \cite{Szabo2020,Westerhout2020}. The neural network quantum states approach was employed for quantum tomography under non-ideal experimental conditions \cite{Torlai2018PRL,Torlai2019,Tiunov2020,Palmieri2020,Danaci2021}. However, how much information is needed for representing a generic quantum state at a given level of accuracy remains an open question \cite{Sciarrino2019,Gebhart2023}.

Instead of characterizing the whole system, one might target only mean values of a set of selected observables, which substantially reduces the required number of measurements. This approach, termed shadow tomography \cite{Aaronson2018}, can also be applied to estimate entanglement entropy of a small subsystem, basically reconstructing its reduced quantum state \cite{Huang2020,Kulik2021}. An alternative method uses random measurements to estimate the second-order R\'enyi entropy of a subsystem \cite{vanEnk2012,Elben2018,Brydges2019,Zoller2022}. However, quantification of entanglement distributed over the whole system lies beyond the scope of such methods.

Entanglement witnessing seems to be a viable alternative to the tomographic methods, when we only aim at distinguishing between entangled and non-entangled states (or between entanglement classes) without quantifying the degree of entanglement and its detailed structure. Nevertheless, the witnessing may still require the full knowledge of the underlying quantum state, as is the case of the positive partial transpose criterion \cite{Horodecki2009}.
The witness cannot be directly measured; however, it can be approximated using a completely positive map \cite{Horodecki2002}, which is equivalent to the full quantum state tomography \cite{Fiurasek2002,Carmeli2016}. Other witnessing methods are based on the minimum local decomposition \cite{Guhne2002,Barbieri2003}, semi-definite programming \cite{Audenaert2006,Jungnitsch2011}, entanglement polytopes \cite{Walter2013}, or correlations in random measurements \cite{Ketterer2019,Knips2020}.
Entanglement witnessing can also be facilitated by using neural networks classifiers \cite{Gao2018,Harney2020,Roik2021}. Despite the success of the entanglement witnessing, it provides only witnesses or lower bounds and often requires some a priori information about the state.

In summary, the connection between the entanglement present in a physical system and the measurements of the correlations of its subsystems is highly nontrivial \cite{Kaszlikowski2008,Micuda2019}.
It seems that full entanglement characterization using single-copy local measurements can only be accomplished with the complete quantum state tomography and, consequently, with exponential scaling of the number of required measurements \cite{Fiurasek2002,Lu2016,Carmeli2016,Yu2020}. The open question remains what one can learn about entanglement from an incomplete observation.

In this work, we use deep neural networks (DNNs) to tackle the problem of entanglement characterization. We develop a method that allows us to quantify the degree of entanglement and quantum correlations in a generic partially mixed state using a set of informationally incomplete measurements. 
The entanglement quantifiers we obtain using DNNs approach are substantially more accurate compared with the values attainable using the state-of-the-art quantum tomography methods.
Also, we demonstrate a measurement-independent quantification of entanglement by developing a deep convolutional network that accepts an arbitrary set of projective measurements without retraining. The DNN-based approaches that we introduce here can be immediately applied for certification and benchmarking of entanglement sources, which we demonstrate by using photonic sources of entangled photons based on spontaneous parametric downconversion and a semiconductor quantum dot.

\begin{figure}[h]
	\centering
	\includegraphics[width=0.8\linewidth]{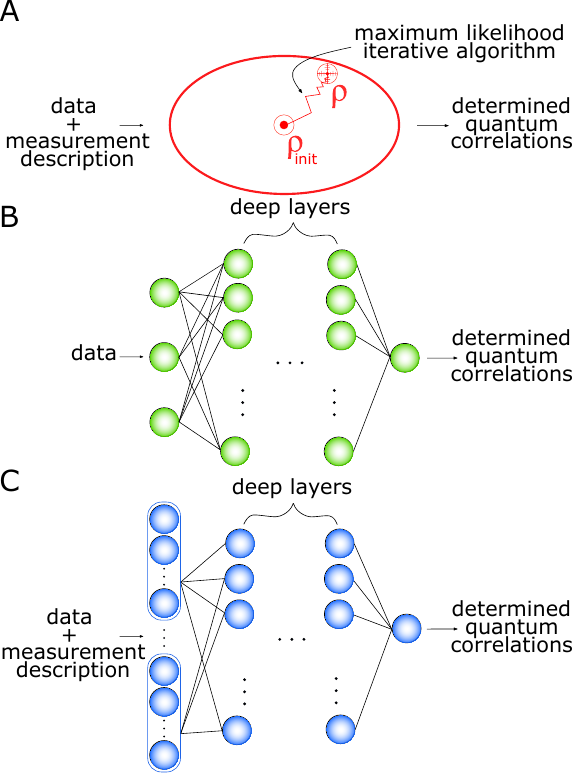}
	\caption{\textbf{Schematics of the three methods we used to infer the quantum correlations.} (A) The maximum likelihood algorithm finds the most likely quantum state $\rho$ based on the measured data and an initial guess $\rho_{\text{init}}$. (B) Green DNN represents a fully connected neural network that infers directly the concurrence and the mutual information from specific measurements (specific measurement projectors), whereas (C) the blue DNN works with an arbitrary measurement projectors. The input for the former are the measured data. The measurement-independent DNN has a first layer convolutional and it inputs both the data and the measurement description.}
	\label{fig:NetFig}
\end{figure}

\begin{figure*}
	\centering
	\includegraphics[width=\textwidth]{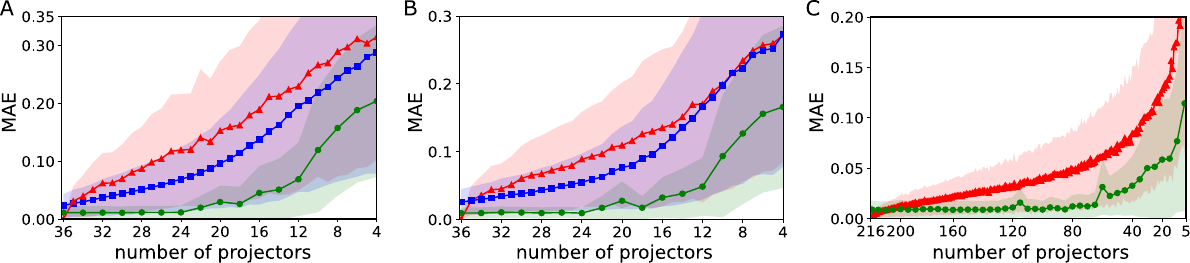}
	\caption{\textbf{Entanglement quantification error for two- and three-qubit systems.}  The mean absolute error (MAE) vs. number of measurement projectors for (A) two-qubit concurrence, (B) two-qubit mutual information, and (C) three-qubit mutual information matrix. Red triangles depict MAE for the MaxLik, blue squares stand for the values of MAE computed from measurement-independent DNN, and finally green circles represent the values of MAE computed from measurement-specific DNNs. The uncertainty regions are depicted in corresponding colors and may overlap. The DNNs outperform the MaxLik approach in terms of entanglement quantification accuracy and its consistency, given by smaller errors and uncertainty intervals, even for substantially incomplete measurements.}
	\label{fig:maeAll}
\end{figure*}

\section{Results}
Even in a well-understood system, such as a pair of qubits, a reliable quantification of entanglement requires full state tomography \cite{Lu2016}. In other words, to infer the degree of entanglement we need to determine the quantum state.  A common approach to implement photonic qubit tomography is to measure the full basis of three Pauli operators. Such a measurement for a two-qubit state consist of $6^2 = 36$  local projectors \cite{Paris2004Springer}.
Omitting randomly some projectors in this measurement scheme decreases the accuracy of the quantum tomography and, consequently, the entanglement evaluation. Here, we show that this problem can be overcome by employing DNNs that allow us to gain knowledge on the degree of entanglement without the need to know the quantum state.

To demonstrate the advantage of the DNN approach we use two quantifiers: the concurrence \cite{Horodecki2009} and the mutual information \cite{Valdez2017} for a two-qubit and a three-qubit system, respectively. The concurrence is widely used in experiments for characterization of entangled photon pair sources. Its value is bounded from below by $0$ for separable states and from above by $1$ for maximally entangled states. On the other hand, the concurrence cannot be easily generalized to higher dimensional quantum systems and systems of more than two parties. Therefore the second quantifier we use is the mutual information, which can be generalized to multipartite systems of qudits and its value reflects the information shared between the parties of a larger system.

We employ three different approaches to determine the concurrence and the mutual information from an incomplete set of data. We show them schematically in Fig.~\ref{fig:NetFig}.  We utilize the maximum likelihood algorithm (MaxLik) (Fig.~\ref{fig:NetFig}(A), red), measurement-specific DNNs, (Fig.~\ref{fig:NetFig}(B), green), and a measurement-independent DNN, (Fig.~\ref{fig:NetFig}(C), blue). The maximum likelihood is an algorithm that finds the quantum state ($\rho$) iteratively, starting from an initial guess ($\rho_{\text{init}}$), which is typically set to maximally mixed state \cite{Hradil2004Springer}. Having at hand the quantum state $\rho$ allows us to quantify the entanglement (see Methods).
In contrast, the approaches based on DNN learn the concurrence and mutual information directly from the measured data. While the measurement-specific DNN is designed for a predefined set of measurement projectors $M_m$, the measurement-independent DNN relaxes the condition on measuring the a priori known projectors and predicts concurrence and mutual information independently on the measurement settings. This approach has a convolutional first layer and it inputs the measured data together with the description of the respective projectors. During the training, the DNNs are provided with the theoretical probabilities $\text{Tr}\{\rho M_m\}$ and, in the case of the measurement-independent DNN, also the description of the measurement $M_m$. 
For the detailed information about the structure of the DNNs, the dataset, and the training procedure see the Methods section.

We compare the three approaches on the basis of how accurately they can infer the concurrence (mutual information) from an incomplete set of data. Here, the MaxLik serves as a benchmark to the other two methods that are DNNs based. We chose to evaluate the performance of all three approaches by computing the mean absolute error (MAE). The MAE is calculated as $\langle|x_i-y_i|\rangle$ with $x_i$ being the true value and $y_i$ the predicted value of the concurrence (mutual information).
To make our comparison universal, the average is taken over a set of states and several combinations of measurement projectors, i.e. a test set. The total number of combinations of $k$ projectors from the maximum of 36 is $\frac{36!}{k!(36-k)!}$. As this number can be excessively high, we randomly selected a smaller subset of combinations. Therefore, to evaluate the performance of a measurement-specific DNN, we train 12 randomly selected networks for each $k$-projector measurement and evaluate the average and standard deviation of their MAEs. For the MaxLik and measurement-independent DNN, the averaging is performed over hundreds of randomly selected measurements.

The performance of the three approaches is presented in Fig. \ref{fig:maeAll}, where we show how MAE depends on the number of measurement projectors we used to obtain the result.
The Figs. \ref{fig:maeAll}(A) and \ref{fig:maeAll}(B) show the MAE for the concurrence and the mutual information, respectively, while the \ref{fig:maeAll}(C) addresses the MAE of the three-qubit mutual information matrix. The MaxLik approach is presented using the red triangles in all panels. For the informationally complete data, i. e. when all 36 projectors are measured, the MaxLik MAE is on the order of $10^{-5}-10^{-4}$. In this scenario MaxLik converges to the true quantum state and the error only reflects the numerical errors caused by the computing precision. As we can see in the Fig. \ref{fig:maeAll}(A) and (B), the MAE of the MaxLik starts increasing if only a few out of the 36 projectors are absent.
In contrast to the MaxLik, DNNs perform well even for a severely reduced number of projectors. The measurement-specific DNNs (shown in green circles) predict the concurrence and mutual information with the MAE of approximately $0.01$ even when only $24$ projectors are used. For the same number of projectors MaxLik MAE is $0.1$. Consequently, measurement-specific DNNs result in a precision that is on average $10$ times higher. If we further reduce the number of projectors, the MAE for the measurement-specific DNNs starts to increase, however, it keeps being substantially smaller compared to the MAE of the MaxLik. Importantly,  the uncertainty region of MAE also remains at least two times smaller (up to ten times while working with more than $18$ projectors). The measurement-independent DNN error is shown in the Fig. \ref{fig:maeAll} using the blue squares. Compared to the performance of the MaxLik, the measurement-independent DNN quantifies the concurrence and the mutual information with a lower MAE, however worse than using the measurement-specific strategy. In practice, one can resource to the measurement-independent DNN for preliminary detecting the entanglement in the system, even changing the measurement on the fly, and improve the entanglement quantification by training a particular measurement-specific DNN later.

\begin{figure}
    \centering
    \includegraphics[width=\linewidth]{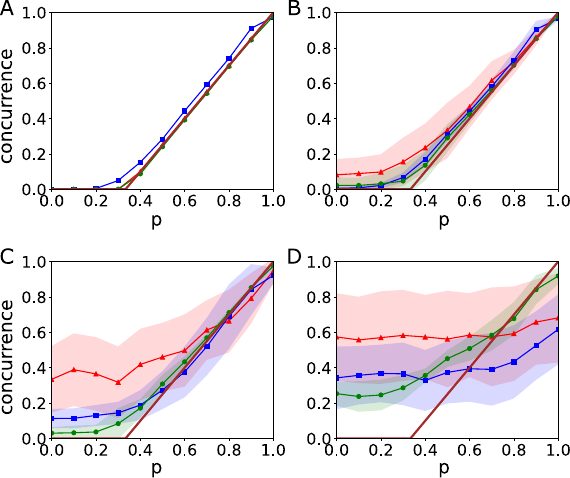}
    \caption{ \textbf{Entanglement quantification error for the Werner state.}The dependence of values of concurrence for the two-qubit Werner state $\rho_W(p)$ characterized by the parameter $p\in [0,1]$.
    Panels (A), (B), (C) and (D) show values of the concurrence determined from 36, 28, 18, and 8 measurement settings, respectively.
    In each panel, the red triangles depict the average values of the concurrence determined by the MaxLik with corresponding uncertainty region, the blue squares stand for the measurement-independent DNN predictions, and the green circles represent predictions given by measurement-specific DNNs. The brown line shows the theoretical values of the concurrence for the Werner state. Both measurement-independent DNN and measurement-specific DNNs outperform the MaxLik in entanglement quantification of the Werner state.}
 \label{fig:concWs}
\end{figure}

To further validate our approach, we compare the values of the concurrence determined by MaxLik, measurement-specific DNNs, and measurement-independent DNN using a state that the network has never seen before, the Werner state~$\rho_W(p) =~ p\rho_{\psi^-}+~\frac{1-p}{4}1$,
where $\rho_{\psi^-}$  is a projector into maximally entangled Bell state spanning the asymmetric subspace of two qubits. The parameter $p$ runs from $0$ (mixed state) to $1$ (maximally entangled state). The concurrence for the Werner state is a piecewise linear function of the parameter $p$ and it takes the exact form $C(\rho_W)=\text{max}\left[0,(3p-1)/2\right]$. The results are shown in the Fig. \ref{fig:concWs}. In the panels (A)-(D) we show the concurrence and the corresponding uncertainty regions for 36, 28, 18, and 8 projective measurements, respectively. 
For 28 and 18 measurement projectors, both the DNN approaches follow the ideal concurrence values while the MaxLik deviates substantially. The measurement-specific DNNs yield nontrivial results even in the case of only 8 measurement projections.

As mentioned previously, the mutual information can be generalized to the systems of more than two qubits. To show that we can also generalize the DNN-based approach to larger quantum systems, we apply our method to a three-qubit system.
In such a system the mutual information matrix has three independent entries $\mathcal{I}\equiv\{\mathcal{I}_{AB},\mathcal{I}_{AC},\mathcal{I}_{BC}\}$, with subscripts referring to the three different ways of partition. To determine all three numbers $\mathcal{I}$ simultaneously, we have to perform a full tomographic measurement on each qubit, which leave us with $6^3 = 216$ projections. Following the procedure introduced for the two-qubit case, we built measurement-specific DNNs, each mapping measurement data to the three-component vector $\mathcal{I}$. Deep layers have the same structure as for quantification of mutual information in the two-qubit case. Final results are shown in the Fig. \ref{fig:maeAll}(C). The MAE of $\mathcal{I}$ is averaged over its three independent elements and over randomly generated quantum states. DNNs predictions are on average akin to the MaxLik ones in the regime close to the complete data. However, with only about third of all projections, measurement-specific DNNs predict the full mutual information matrix on average with five times smaller error than the MaxLik.

Our approach needs modest computational resources. Particularly, the 2-qubit and 3-qubit measurement-specific networks (for 1/4 of all Pauli projectors compared to the complete measurement) have approximately 37 thousand and 42 thousand parameters, respectively. The optimal performance of networks for 3 qubits does not require substantially more parameters than for 2-qubit networks. We further verified this optimistic scaling by training 4-qubit and 5-qubit measurement-specific networks (for 1/4 of all possible Pauli projectors in each case). These networks require 69 thousand and 231 thousand parameters, respectively, and outperform the MaxLik even more than 2-qubit and 3-qubit networks, see Table~\ref{tab:results}. Namely, the measurement specific networks reach 2.2, 3.0, 3.8, and 4.3 times lower MAE of mutual information matrix than the MaxLik for 2, 3, 4, and 5 qubits, respectively. Based on this finding, we expect that by keeping constant the ratio of the MaxLik accuracy and the DNN accuracy the required fraction of the projectors with respect to the full tomography will decrease.

\begin{figure}
	\centering
    \includegraphics[width=\linewidth]{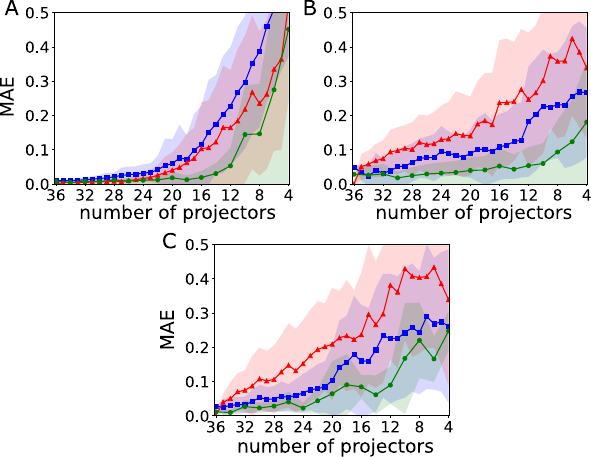}
    \caption{\textbf{Performance of MaxLik and DNN based approaches for an experimental data sets.} We show the dependence of the MAE on the number of projectors. for the spontaneous parametric downconversion (SPDC) sources, panels (A) and (B), and semiconductor quantum dot source, panel (C). The concurrence of experimentally prepared quantum states was determined from the full MaxLik tomography to (A) $0.985 \pm 0.001$, (B) $0.201 \pm 0.002$ and (C) $0.18 \pm 0.01$. The MAE for the measurement-specific DNNs is depicted in a green circles, measurement-independent DNN in a blue squares and the MaxLik approach in a red triangles.
    }
 \label{fig:concExp}
\end{figure}

\begin{table}[h!]
\centering
\begin{tabular}{c|cc|c}
\multirow{2}{*}{\begin{tabular}[c]{@{}c@{}}number\\ of qubits\end{tabular}} &
  \multicolumn{2}{c|}{MAE} &
  \multirow{2}{*}{\begin{tabular}[c]{@{}c@{}}ratio of MaxLik\\ and DNN MAEs\end{tabular}} \\ \cline{2-3}
  & \multicolumn{1}{c|}{MaxLik}  & DNN     &        \\ \hline
2 & \multicolumn{1}{c|}{ 0.20 $\pm$ 0.16 } & { 0.09 $\pm$ 0.09 } & {2.2} \\
3 & \multicolumn{1}{c|}{0.068 $\pm$ 0.055}   & {0.023 $\pm$ 0.020}   & {3.0}       \\
4 & \multicolumn{1}{c|}{0.019 $\pm$ 0.014}   & {0.005 $\pm$ 0.001}   & {3.8}       \\
5 & \multicolumn{1}{c|}{0.039 $\pm$ 0.032 }  & {0.009 $\pm$ 0.001}   & {4.3}     
\end{tabular}
\caption{The summary of the mutual information quantification from incomplete measurements consisting of 1/4 of all possible Pauli projectors in each case. The MaxLik and the measurement-specific DNNs are compared up to 5-qubit quantum systems. The ratio of the mean absolute errors (MAEs) of the methods shows increasingly improving the performance of the DNN approach for entanglement quantification in higher dimensional systems.}
\label{tab:results}
\end{table}

Finally, we demonstrate the performance of DNN-based entanglement quantification using experimental data acquired under non-ideal conditions and with limited statistical sampling. We study two distinct entanglement sources. The first one is based on continuously pumped spontaneous parametric downconversion. The photon pair generation process is inherently random and the resulting entangled state depends on the choice of the temporal coincidence window and other experimental conditions such as background noise. Adjusting of the experimental parameters affects the degree of entanglement in the produced state. We quantify concurrence using the DNNs and the MaxLik approach for various experimental settings ranging from the maximally entangled singlet Bell state to a noisy state with a negligible concurrence. Fig.~\ref{fig:concExp}(A) and (B) show the results for an almost pure entangled state and a partially mixed state with the concurrence of  $0.985\pm0.001$ and $0.201\pm0.002$, respectively. In both cases, DNN approaches outperform the MaxLik approach. The measurement-specific DNNs remain very accurate (MAE $<0.04$) all the way down to $14$ projections. Even the measurement-independent DNN outperform the MaxLik in the generic case of partially mixed state for any number of measurement projectors.
The maximally entangled state represents the only case where the MaxLik performs slightly better than the measurement-independent DNN (but worse than measurement-specific DNNs). This behavior results from high purity and sparsity of the state and, consequently, from the sparsity of the measurement data. When randomly selecting a subset of projectors, there is a high possibility of having a majority or even all the measurements with a negligible number of detection counts. It seems that the predictive strength of the measurement-independent DNN is limited for such a scenario. However, the MaxLik approach is biased towards pure states in the case of heavily undersampled data \cite{Guehne2015,Silva2017}, and the positivity constraint tends to a sparse (low-rank) states \cite{Kalev2015}.
This bias artificially increases the resulting concurrence and reduces its error.

The second experimental system consists of a semiconductor quantum dot resonantly pumped by picosecond pulses. The biexciton-exciton cascade emission produces pairs of photons in a partially polarization entangled state. The degree of entanglement is reduced by the  presence of the fine-structure splitting reaching the concurrence of $0.18\pm 0.01$. Fig. \ref{fig:concExp}(C) shows the MAE for such a mixed quantum state. As for the source based on spontaneous downconversion, both DNNs approaches beat on average the MaxLik method in accuracy. Let us point out that the DNN based approaches were trained to predict quantum correlations from the theoretical probabilities computed from the ideal quantum states and measurement. The Fig. \ref{fig:concWs} thus demonstrates the robustness of our approaches to noisy experimental data.

\section{Conclusion}

We demonstrated that by exploiting novel methods of neural networks and deep learning we can outperform the traditional and commonly used techniques for quantification of quantum correlations such as state tomography. For the systems of two qubits, we built two different neural-network-based approaches, namely measurement-specific and measurement-independent DNNs. Both approaches predict concurrence and mutual information from data with a higher accuracy than the commonly used quantum state tomography. The best performing approach is the measurement-specific DNNs, which are trained to predict the concurrence or mutual information from a fixed set of projectors. Furthermore, we generalized to the system of three qubits, where we show that the measurement-specific DNNs represent a more accurate method to quantify the mutual information matrix than the maximum likelihood one. We demonstrated the feasibility of the measurement-specific DNNs training up to five qubits. Our approaches not only benefit from high accuracy when working with fewer measurement projectors, but also are substantially faster compared to the standard tomography-based methods. Furthermore, we demonstrate the robustness of our approach using two experimental systems: a nonlinear parametric process and a semiconductor quantum dot. The DNN approaches can be further studied and modified to adaptively find a minimal set of projectors that infer the entanglement accurately.

\section{Methods}\label{sec:Methods}
\footnotesize

\noindent\textbf{Quantifying quantum correlations.}  To quantify the quantum correlation we use the concurrence and the mutual information, for two qubit and three-qubit case, respectively. The concurrence is a two-qubit monotone entanglement measure \cite{Horodecki2009} widely used for the characterization of bipartite entanglement commonly present in sources of entangled photon pairs. Knowing the quantum state the concurrence is defined as
\begin{equation}\label{eq:concur}
\mathcal{C}\left(\rho \right) \equiv \text{max}\left\{0,\lambda_1-\lambda_2-\lambda_3-\lambda_4\right\},\\
\end{equation}
with $\lambda_1,\dots,\lambda_4$ being the eigenvalues (sorted in decreasing order) of the Hermitian matrix $T = \sqrt{\sqrt{\rho}\tilde{\rho}\sqrt{\rho}}$, here $\tilde{\rho}=\sigma_y \otimes \sigma_y \rho^* \sigma_y \otimes \sigma_y$ where $\rho^*$ standing for complex conjugate and $\sigma_y$ is one of the Pauli matrices represented in a computational basis as $\sigma_y=i\left(|1\rangle\langle 0|-|0\rangle\langle 1|\right)$.
For an arbitrary mixed state, the value of concurrence is saturated from below by $0$ \cite{Horodecki2009} for the separable states $\rho_{AB} = \sum_{i}\gamma_{i}\rho_{A}^{i}\otimes\rho_{B}^{i}$ and from above by $1$ for the maximally entangled states of two-qubits.

Mutual information is a quantum correlation measure commonly used in quantum cryptography or for quantifying complexity in many-body systems. For an $n$--qubit quantum system, mutual information matrix reads
\begin{equation}\label{eq:MutInf}
\mathcal{I}_{ij}=\frac{1}{2}\left(\mathcal{S}\left(\rho_{i}\right)+\mathcal{S}\left(\rho_{j}\right)-\mathcal{S}\left(\rho_{ij}\right)\right),
\end{equation}
and is constructed from the one and two point von Neumann entropies \cite{Valdez2017}, $\mathcal{S}\left(\rho_{i}\right)=-\text{Tr}\{\rho_i\log_d\rho_i\}$, $\mathcal{S}\left(\rho_{ij}\right)=-\text{Tr}\{\rho_{ij}\log_d\rho_{ij}\}$, with $\rho_{i}$ and $\rho_{ij}$ standing for reduced density matrices, $\rho_{i}=\text{Tr}_{k\ne i}\{\rho\}$, $\rho_{ij}=\text{Tr}_{k\ne ij}\{\rho\}$ respectively.\\

\noindent\textbf{Quantum state tomography.} Quantum state tomography is a method to solve the inverse problem of reconstruction of an unknown quantum state. It uses the set of measurement operators (projectors) and relative frequencies $\{f_i\}$ acquired in a measurement. To obtain an informationally complete measurement we need the relative frequencies for at least $D^2-1$ independent projectors $\{M_i\}_{i=0}^N$.
Quantum state is reconstructed by maximizing the log-likelihood functional $\mathcal{L}(\rho) \propto \sum_{j=1}^{N} f_j \log p_j(\rho)$, which can be written \cite{Hradil1997,Jezek2003,Hradil2004Springer} as the iterative map $\rho^{(k+1)} \leftarrow \mu_{k} R\rho^{(k)} R$,
where $\mu$ is the normalization constant and $R$ an operator defined as $R=\sum_if_i/p_iM_i$. Here, $f_i$ are the measured frequencies and $p_i$ are the theoretical probabilities given by the Born's rule $p_i=\text{Tr}\{\rho M_i\}$. In the main text of this paper, we address how the measurement being incomplete affects the quantification accuracy of the concurrence and the mutual information. In such a case, the closure relation, $\sum_i M_i = 1$,  is no longer fulfilled. The optimal strategy is to map the set of projectors $\{M_i\}$ into a new set $\{M_i'\}$ via $M_i' = G^{-1/2}M_iG^{-1/2}$ with $G=\sum_i M_i$. One can easily check that the $\{M_i'\}$ now fulfill the completeness relation, $\sum_i M_i' = 1$.  The iterative map then updates to
\begin{equation}\label{eq:ml}
\rho^{(k+1)} \leftarrow \mu_{k} G^{-1/2}R\rho^{(k)} RG^{-1/2},
\end{equation}
and represents a procedure that we follow in the main text. We consider measurement settings to be the Pauli projectors , i.e. projectors into eigenstates of Pauli operators $\{\sigma_x,\sigma_y,\sigma_z\}$. The MaxLik estimator $\rho^{\text{MaxLik}}$ is defined as a fixed point of the iterative map (\ref{eq:ml}).
The iteration process starts from the completely mixed state $\rho_{\text{init}} = 1/D$ and is stopped when the Hilbert-Schmidt distance between the subsequent iterations reaches $10^{-16}$. 
In the case of encoding qubit states into the polarization degrees of freedom, Pauli measurement consist of projectors onto three mutually unbiased basis sets $\{|H\rangle\langle H|, |V\rangle\langle V|,$ $|D\rangle\langle D|, |A\rangle\langle A|,$ $|R\rangle\langle R|, |L\rangle\langle L|\}$.

There are other methods for quantum state tomography such as maximum-likelihood maximum-entropy \cite{Teo2011}, semidefinite programming \cite{Vandenberghe1996}, or compressed sensing \cite{Gross2010}. These methods and their comparison to MaxLik and DNNs are presented in the Supplementary Materials.\\

\noindent\textbf{Deep neural networks methods.} Neural networks are machine learning models that learn to perform tasks by analyzing data. A DNN model consists of multiple layers of interconnected artificial neurons and acts as a highly nonlinear transformation parametrized by a large number of trainable parameters \cite{LeCun2015}.
DNNs possess the ability to generalize from learning stage, i.e. once trained they can perform surprisingly well even for inputs that were not observed during the learning stage.
The basic principles of DNNs operation are well known, but the full span of their generalization ability is the subject of current research \cite{Belkin2019,Kawaguchi2022}.
In science and technology, neural networks have been successfully applied to a wide range of problems, including predicting the behavior of complex systems and analyzing large datasets from experiments and simulations \cite{Carleo2019RMP,Neupert2021}.

Let us first consider the DNN quantification of entanglement in a two-qubit system. The measurement-specific DNNs are fully connected networks. The network has seven fully connected layers with a few dozens of thousands trainable parameters in total. The exact number of the free parameters differs between the networks that have different length of the input vector, dependent on the number of projectors measured. We trained 193 measurement-specific DNNs (12 per point except of full 36 projectors) with varying length of the input layer, starting with the full 36 input neurons down to 4 (with increment of 2).

We construct a set of quantum states as follows: We generate $10^6$ random quantum states $\rho$ of which $4/5$ are randomly distributed according to the Bures measure induced by the Bures metric \cite{Bures1969},
\begin{equation}
\label{eq:Bures}
    \rho = \frac{\left(1+U^{\dag}\right)GG^{\dag}\left(1+U\right)}{\text{Tr}{\left(1+U^{\dag}\right)GG^{\dag}\left(1+U\right)}}.
\end{equation}
To achieve this, we generate a Ginibre matrix $G$ with complex entries sampled from the standard normal distribution, $G_{ij} \sim \mathcal{N}\left(0,1\right) + i \mathcal{N}\left(0,1\right)$, together with a random unitary $U$ distributed according to the Haar measure \cite{Mezzadri2007}.
The remaining $1/5$ of the set consists of random Haar pure states mixed with white noise.
The generation of the set aims at the most uniform and broadest coverage of partially mixed quantum states.
The set of quantum states is randomly shuffled and split to two parts, i.e. the training and validation sets containing 800,000 and 200,000 samples, respectively. The test set has 5,000 states generated according to the Eq.(\ref{eq:Bures}).

For the quantum states we prepare the corresponding datasets by computing the probability distribution with elements $\textbf{p}_i\equiv\text{Tr}\{\rho M_i\}$ and evaluate the quantum correlation measure (concurrence or mutual information) using Eqs.(\ref{eq:concur} and \ref{eq:MutInf}).
We trained the measurement-specific DNNs to predict the quantum correlations from the probability distribution $\textbf{p}$. The training and validation datasets have the following structures
\begin{equation}
\begin{split}
\mathcal{D}^{\text{input}} &= \{\text{Tr}\{\rho M_1\},\dots,\text{Tr}\{\rho M_{36}\}\},\\
\mathcal{D}^{\text{output}} &= \{\mathcal{Q}(\rho)\}, \\
\end{split}
\end{equation}
where the length of the input vector $\mathcal{D}^{\text{input}}$ is different for various measurement-specific neural networks, ranging from full $36$ projectors down to $4$. The output $\mathcal{Q(\rho)}$ stands for either the  concurrence or the mutual information.

We achieve the learning of the neural networks by backpropagating the error through the use of chain rule of derivation. It minimizes the loss function defined as the mean absolute difference between the true values of the quantum correlations measure $\mathcal{Q}_{\text{true}}$ and the values $\mathcal{Q}_{\text{predicted}}^{\theta}$ predicted by the networks. The loss function thus takes a form
\begin{equation}
\label{eq:loss}
\mathcal{L}= \left \langle \left | \mathcal{Q}_{\text{true}} - \mathcal{Q}_{\text{predicted}}^{\theta} \right | \right \rangle,
\end{equation}
and the minimum is found by minimizing the $\mathcal{L}$ over all components of a training dataset to update weights and biases $\{\theta\}$ using the Nesterov-accelerated adaptive moment estimation (NAdam) algorithm. 
At the step $t$, the NAdam procedure updates parameters
\begin{eqnarray}
\theta_t \leftarrow \theta_{t-1} - \eta \frac{\bar{\mathbf{m}}_t}{\sqrt{\hat{\mathbf{n}}_t}+\epsilon}, 
\end{eqnarray}
with
\begin{eqnarray}
\mathbf{g}_t \leftarrow \nabla_{ \theta_{t-1}} \mathcal{L}\left(\theta_{t-1}\right), \nonumber \\
\mathbf{\hat{g}} \leftarrow \frac{\mathbf{g}_t}{1-\prod_{j=1}^{t}\mu_j}, \nonumber \\
\mathbf{m}_t \leftarrow \mu \mathbf{m}_{t-1}+(1-\mu)\mathbf{g}_t,  \nonumber \\
\mathbf{\hat{m}}_t \leftarrow \frac{\mathbf{m}_t}{1-\prod_{j=1}^{t+1}\mu_j},  \\
\mathbf{n}_t \leftarrow \nu \mathbf{n}_{t-1}+(1-\nu)\mathbf{g}_t^2, \nonumber\\
\mathbf{\hat{n}}_t \leftarrow \frac{\mathbf{n}_t}{1-\nu^t}, \nonumber \\
\mathbf{\bar{m}}_t \leftarrow (1-\mu_t) \mathbf{\hat{g}}_t +\mu_{t+1}\mathbf{\hat{m}}_t. \nonumber 
\end{eqnarray}
The parameter $\eta$ represents the learning rate, parameter $\mu$ represents the exponential decay rate for the first moment estimates $\mathbf{\hat{m}}$, the parameter $\nu$ is the exponential decay rate for the weighted norm $\mathbf{g}_t^2$ and $\epsilon$ is a parameter that ensures the numerical stability of the NAdam optimization procedure. In our work, we set the numerical values of parameters $\{\eta,\mu,\nu,\epsilon\}$ to $\{0.001,0.9,0.999,10^{-7}\}$.
The training takes over 2,000 epochs with data further divided into $100$ batches to optimize the learning time and accuracy of the predictions on the validation dataset.
All above is implemented using Keras and Tensorflow libraries for Python.

The measurement independent DNN is a generalization to the measurement-specific DNNs, and therefore it consists of single network that predicts the concurrence and the mutual information from any set of projectors that we chose to work with. This functionality is accomplished by a restructuralization of the input layer that inputs not only the vector of probabilities $\textbf{p}$ but also the description of the measurement $\{M_i\}$ itself,
\begin{equation}
\mathcal{D}^{\text{input}} = \{M_1,\text{Tr}\{\rho M_1\},\dots,M_{36},\text{Tr}\{\rho M_{36}\}\}.
\end{equation}
The kernel of the first convolutional layer has a stride length equal to the length of the pair $\{M_i,\text{Tr}\{\rho M_i\}\}$ to prevent the network to see crosstalk between the adjacent input pairs. Each projector $M_i$ is vectorized using $d^2$ trace orthonormal basis operators $\{\Gamma_i\hspace{0.1cm}|\hspace{0.1cm}\Gamma_i\geq 0,\hspace{0.1cm} \hspace{0.1cm} \text{Tr}\{\Gamma_{i}\Gamma_{j}\}=\delta_{ij}\hspace{0.1cm}\forall i,j\}$. For incomplete measurements containing less that 36 projectors, we set the values of missing measurement probabilities and projectors to zero.

For three-qubit, four-qubit, and five-qubit systems, the structure, the loss function, and the optimization procedure of the measurement-specific DNNs are the same as for the systems of two qubits.
We trained $44$ measurement-specific DNNs for three qubits. The length of the input vector is different for each measurement-specific neural network, ranging from $6^3 = 216$ down to $5$.
We also trained two specific networks for four-qubit and five-qubit systems with 325 and 1944 input measurements, respectively.
The training and validation datasets are divided in a ratio $4:1$. They contain 100,000 measurement probability distributions (input) and values of the mutual information (output) computed from 100,000 quantum states generated using the same process as for two-qubit states.
The number of training points in the dataset is lower than in the two-qubit case due to memory limitations. For this reason, we adopted the incremental learning method \cite{Carpenter1991}. After the loss function on the validation dataset reaches minimum which is not updated in the next 200 epochs, the training is stopped and the best model is saved. Next, we generate different 100,000 data points and continue training. The test set consists of 500 states generated according to Eq.(\ref{eq:Bures}).

The complexity of the developed DNNs is rather low, and their scaling to higher dimensional systems is feasible. The largest network presented (two-qubit device-independent DNN) has almost 460 thousand trainable parameters. The 5-qubit DNN has slightly more than 230 thousand trainable parameters. Its training on 2 million data samples takes 45 hours on a single consumer-grade GPU.
With larger computational resources (available today) we believe that training the networks for entanglement quantification in systems with dozens of qubits should be feasible.
The conventional methods, such as MaxLik, are also computationally demanding and have to be evaluated for every new data. In contrast, our approach is computationally demanding only in the training stage. The forward evaluation (from data to entanglement) is computationally easy. Specifically, the DNN entanglement quantification is on average four orders of magnitude faster than the MaxLik and two orders of magnitude faster than the SDP.\\

\noindent\textbf{Experiment.} The spontaneous parametric downconversion source consists of a beta barium borate (BBO) crystal cut for type-II colinear generation of two correlated orthogonally polarized photons with the central wavelength of 810 nm. The BBO crystal was pumped by a continuous laser. An entangled singlet polarization state was conditionally generated by interfering the correlated photons at a balanced beamsplitter.
To achieve the complete set of data we performed the full quantum state tomography. This was performed by measuring all 36 projective measurements as combinations of local projections to horizontal, vertical, diagonal, anti-diagonal, right-hand and left-hand circular polarizations. The polarization analyzer consists of a sequence of half-wave and quarter-wave plates followed by a polarizer, single-mode fiber coupling, and a single-photon detector. The detection events from the two detectors were taken in coincidence basis.
To obtain the data sets where the entanglement was reduced by noise one of the pair photons was propagated through a noisy channel. The noise was implemented by injecting a weak classical signal from an attenuated laser diode.
The concurrence of the entangled state reached 0.98 for a short coincidence window and no injected noise. However, for larger coincidence windows and higher levels of injected noise, the concurrence of the detected state decreased. The experimental data for the entangled states with the concurrence of $0.985\pm0.001$ and $0.201\pm0.002$ used in this work were acquired in \cite{Straka2015}.

Semiconductor quantum dot source consists of a quantum dot embedded in a circular Bragg grating cavity \cite{Kartik_BE} that enables high photon collection efficiency. The quantum dot was excited via two-photon resonant excitation of the biexciton \cite{twophoton}. The excitation pulses were derived from a pulsed 80MHz repetition rate Ti:Sapphire laser. The laser scattering was spectrally filtered, and the exciton and biexciton emission were separated ahead of single-mode fiber coupling. The polarization state of the generated entangled state was analyzed using two polarization analyzers in the process of full quantum state tomography in the same way as it was performed for the parametric downconversion source. The observable degree of entanglement was predominantly limited by the non-zero fine structure splitting.

\normalsize

\section*{Data availability}

Data underlying the results presented in this paper are publicly available in Ref.~\cite{KoutnyGitHub2022}.

\section*{Funding}

The Czech Science Foundation (grant No.~21-18545S);
The Ministry of Education, Youth and Sports of the Czech Republic (grant No.~8C18002);
European Union's Horizon 2020 (2014--2020) research and innovation framework programme (project HYPER-U-P-S);
D.K. acknowledges support by Palack\'y University (grants IGA-PrF-2021-002 and IGA-PrF-2022-001);
A.P. would like to acknowledge Swedish Research Council. Project  HYPER-U-P-S  has  received  funding  from  the QuantERA ERA-NET  Cofund in  Quantum  Technologies  implemented  within  the  European  Union’s Horizon  2020  Programme.
L. G. was supported by the Knut \& Alice Wallenberg Foundation (through the Wallenberg Centre for Quantum Technology (WACQT)).
S.H. acknowledges financial support by the State of Bavaria.

\section*{Acknowledgments}
We thank J. Fiur\'a\v{s}ek for fruitful discussion.
We acknowledge the use of cluster computing resources provided by the Department of Optics, Palacký University Olomouc. We thank J. Provazník for maintaining the cluster and providing support.

\section*{Author Contributions}

D.K. and M.J. provided the theoretical analysis. D.K. performed numerical simulations and experimental data analysis. L.G. and A.P. performed the measurements on quantum dot source. M.M.D., S.H., and C.S. fabricated the sample. M.J. initiated and coordinated the project. D.K., A.P. and M.J. wrote the manuscript. All authors were involved in revising the manuscript.

\section*{Disclosures}

The authors declare no competing interests.

\bibliographystyle{naturemag}
\bibliography{paperConc}



\section*{Supplementary material}

# SUPPLEMENTARY MATERIALS

Definitions of MLME and SDP tomographic methods and their comparison with MaxLik and DNNs:

In the main text we compare the direct DNN approach to indirect quantum state reconstruction. We use the MaxLik for the quantum state reconstruction and we describe it in the Methods. Here, we review other approaches for quantum state reconstruction, namely maximum-likelihood maximum-entropy (MLME) and semidefinite programming (SDP). Furthermore, we discuss the convergence properties in the case of incomplete measurements and compare the performance of all methods for ideal probabilities and finite sampled data.

The MLME [83] is based on maximization of the log-likelihood functional together with the von Neumann entropy $\mathcal{S}(\rho) = -\text{Tr}\{\rho \log \rho\}$. The total functional reads $\mathcal{F}(\rho, \lambda) \propto \mathcal{L}(\rho) + \lambda \mathcal{S}(\rho)$. In a similar way as for the log-likelihood functional, one can derive from $\mathcal{F}$ an iterative map $\rho^{(k+1)} \leftarrow \mu_k T \rho^{(k)} T$. The operator $T$ takes the form $R - \lambda(\log \rho_k - \text{Tr}\{\rho_k \log \rho_k\})$, with $R$ being the same as in Eq. (3) of the main text, $R = \sum_i f_i/p_i M_i$, and $\mu_k$ is the normalization constant ensuring the unit trace of the estimator. The MaxLik and the MLME are both constrained to positive estimates, $\rho \geq 0$.

In the SDP [30,84] the estimator is a solution to the constrained optimization problem, $\rho_{\text{est}} = \min_\rho ||\mathcal{A}(\rho) - \mathbf{p}||$, such that $\rho \geq 0$. The operator $\mathcal{A}$ represents the action of projectors $M_i$ on the quantum state according to the Born's rule $\text{Tr}\{\rho M_i\}$ and $\mathbf{p}$ is the vector of probabilities. We considered two widely used norms $||\cdot||$ as a cost function, namely the $l_1$ norm defined on elements of a vector space $\mathbf{x}$ as $||\mathbf{x}||_{l_1} = \sum_i |x_i|$ and the $l_2$ norm that acts on vectors as $||\mathbf{x}||_{l_2} = \sqrt{\sum_i |x_i|^2}$.

Let us comment on how the introduced quantum state reconstruction methods cope with incomplete data. The MaxLik converges to an estimate within a plateau, which is given by input data and an initial iteration. In our case, we start from a maximally mixed state, which spans the whole Hilbert space and allows the iteration process to go "anywhere". It was shown that the MaxLik method represents an unbiased estimator except for very small amount of measurement data and states close to pure states [78,79]. The MLME converges to a state with maximum entropy within a plateau of states given by the measurement data likelihood. Here, we also start from a maximally mixed state for the same reason as given for the MaxLik. The SDP solves a linear positive problem by the primal-dual interior-point method. Based on what cost function is used, the SDP tends to a more or less sparse solution. The sparse (low entropy) solution is usually connected to the L1 norm. Finally, let us point out that minimizing any cost function subjected to the positivity constraint is a compressed sensing protocol [80].

Here, we compare our novel DNN approach to quantification of entanglement with the four quantum state reconstruction methods we listed above. We compute the MAE between true and inferred values of the concurrence in the same way as in the main text. The results are plotted in Fig. 1 for a two-qubit system and the ideal measurement probabilities (infinite sampled data). The DNNs outperform significantly the other methods. On average, the MaxLik approach quantifies the entanglement with either better or similar accuracy when compared to the MLME and SDP. The only exceptions are measurement with very low number of projectors, where the SDP methods slightly outperform the MaxLik. However, considering their large uncertainty regions, all four quantum-state reconstruction methods perform on par. These results show that (*i*) the DNN approach is superior, and (*ii*) the MaxLik can be used as an representative example of quantum-state reconstruction method.

We compare the methods also for noisy data (finite sampled), see Fig. 2. The results remain on par with those obtained in the case of the ideal measurement probabilities (infinite sampled data). Moreover, for a small number of the total detections (e.g. one thousand samples), the measurement-specific DNNs outperform the other methods even for the full 36-projector measurement.

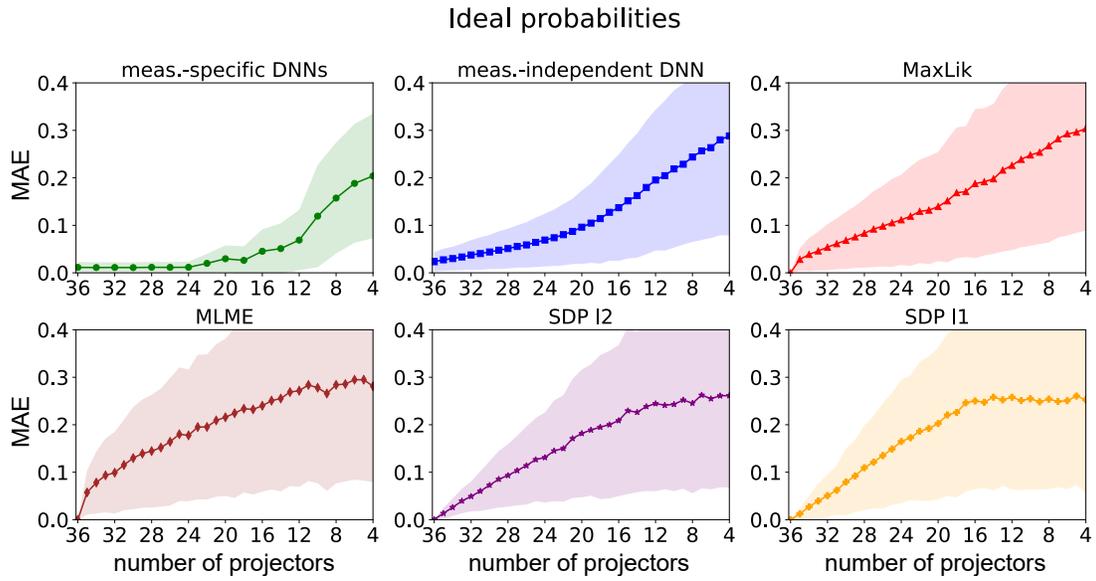

Figure 1. Comparison of mean absolute error (MAE) of concurrence for different tested methods. The input simulated data are the ideal measurement probabilities. Green circles represent the values of MAE computed from measurement-specific DNNs, MAE for the measurement-independent DNN is depicted in blue squares, Maxlik in red triangles, MLME algorithm in brown diamonds, the SDP with $l_2$ norm in violet stars and the SDP with $l_1$ norm in orange crosses.

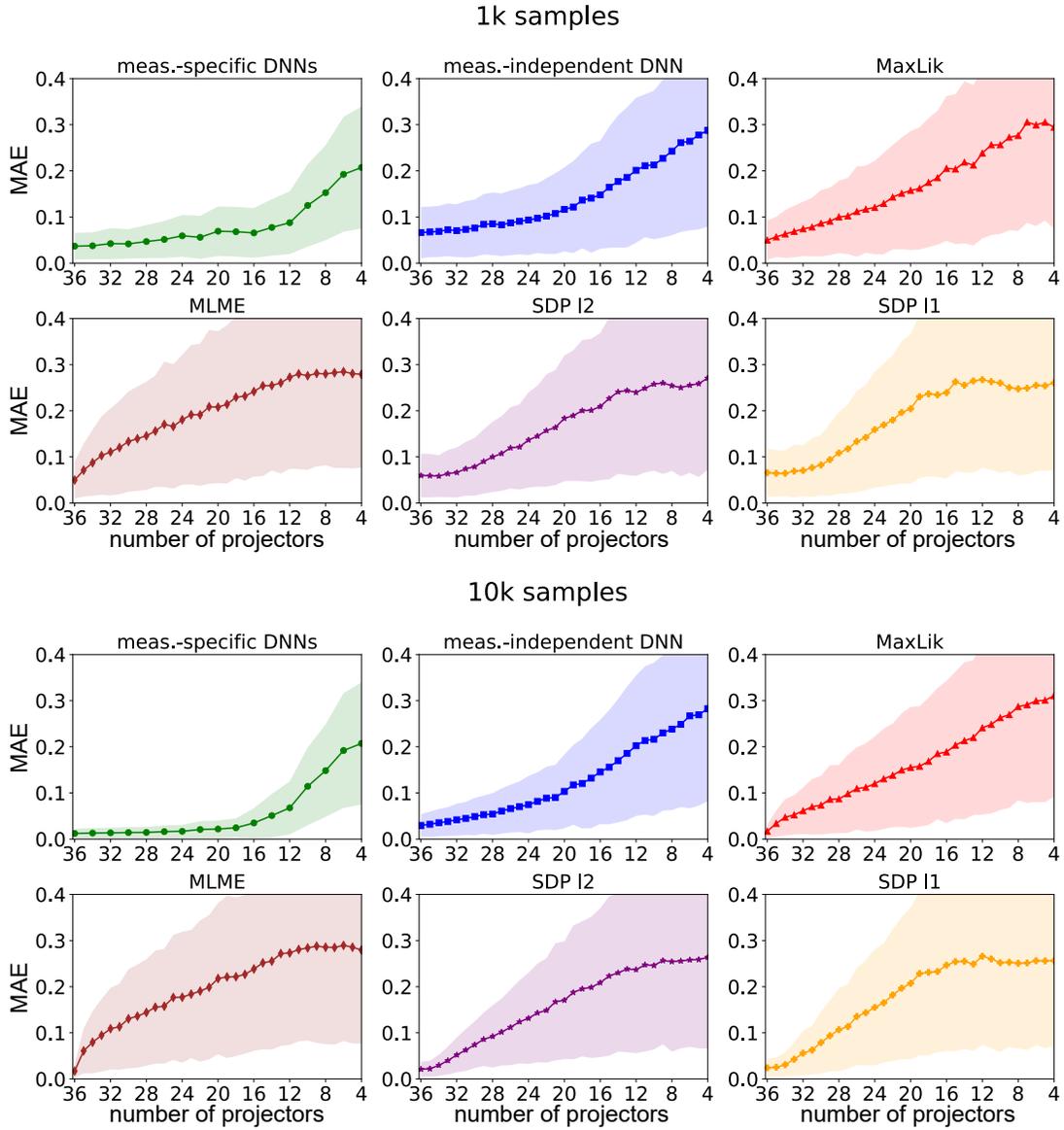

Figure 2. Comparison of mean absolute error (MAE) of concurrence for different tested methods. The input simulated data are finite sampled; two cases are shown with $10^3$ and $10^4$ total detections, respectively. Green circles represent the values of MAE computed from measurement-specific DNNs, MAE for the measurement-independent DNN is depicted in blue squares, Maxlik in red triangles, MLME algorithm in brown diamonds, the SDP with $l_2$ norm in violet stars and the SDP with $l_1$ norm in orange crosses. Each point is averaged over 1000 randomly generated quantum states and uncertainty regions are depicted in corresponding colors.




\end{document}